\documentclass[9pt,twocolumn,twoside]{osajnl}
\graphicspath{{figures/}}
\journal{josab} 

\setboolean{shortarticle}{false} 

\title{Approximations used in the analysis of signals in pump-probe spectroscopy}

\author[1,*]{Peter Karlsen}
\author[1]{Euan Hendry}

\affil[1]{School of Physics, University of Exeter, Stocker Road, EX4 4QL, United Kingdom}

\affil[*]{Corresponding author: Peterkarlsen88@gmail.com}

\ociscodes{(070.4790) Spectrum analysis; (160.4236) Nanomaterials; (260.5150) Photoconductivity; (300.6495) Spectroscopy, terahertz; (300.6500) Spectroscopy, ultrafast; (310.6860) Thin films, optical properties.}

\begin{abstract}
In pump-probe spectroscopy, one often needs to analyse the transmission or reflection of electromagnetic waves through optically pumped media. Here, it is common practice to approximate the analysis in order to extract non-equilibrium values for the optical constants of the media. Concentrating on optical pump-THz probe spectroscopy, and using the transfer matrix approach, we present a general method for evaluating the applicability of the most common approximations. Somewhat surprisingly, we find that these approximations are truly valid only in extreme cases where the optical thickness of the sample is several orders of magnitude smaller or larger than the probe wavelength.
\end{abstract}

\setboolean{displaycopyright}{true}

\begin{document}

\maketitle

Optical pump - THz probe time-domain spectroscopy (OPTP) is a powerful tool for studying the ultrafast charge-carrier dynamics of materials on a femtosecond to nanosecond time-scale, and has successfully been applied to a range of different materials including semiconductors, carbon nanotubes, nanowires, and polymers \cite{Ulbricht2011,Jepsen2011,Joyce2016,Schmuttenmaer2004}. The power of this method lies in the extraction of the complex material parameters (represented by a complex permittivity, refractive index or conductivity) from photoinduced changes to the THz transmission/reflection of a sample \cite{Duvillaret1996a,Nienhuys2005,Kuzel2007}. This approach has allowed the study of a wealth of photo-species and phenomena, including exciton dynamics \cite{Bergren2016}, plasmon formation \cite{Huber2005}, coulomb screening \cite{Jensen2014a} and carrier scattering \cite{Beard2000}.

Extracting the exact, non-equilibrium material parameters of a sample in an OPTP experiment is not trivial, requiring full wave solutions to Maxwell's equations, which must be numerically solved \cite{Hempel2017,Hempel2016}. However, depending on the geometry and dimensions of the sample, in some cases a set of simplified equations may be used to obtain approximate material parameters of the photoexcited sample, as shown in \cite{Kuzel2007}. The most commonly used approaches assume either a sample which is thin or thick relative to the wavelength of the incident THz radiation, and have been used by numerous groups throughout the literature for a variety of thin- \cite{Cunningham2008,Zajac2014,Xiao2015,Ziwritsch2016,Strothkamper2012,Nemec2009,Nemec2008,Strothkamper2013,Nemec2013,Liu2012,HynekNemec2008,Cunningham2013,Jnawali2013a,Terashige2015,Yettapu2016,Lui2001,Prasankumar2005,Minami2015,Hempel2017,Cooke2012,Cooke2012a,Valverde-Chavez2015,Nienhuys2005,Nienhuys2005a,Alberding2016,Xing2017,Jin2014,Joyce2016,Nemec2015,Beaudoin2014,Lu2013,Tang2012,Jnawali2015,Pijpers2010,Xu2009,Kuzel2007,Jensen2014a,Zhang2017,Petersen2017} and thick-films \cite{Hendry2005,Hendry2006a,Nienhuys2005,Shan2003,Knoesel2004,Dakovski2007,Kunneman2013,Ulbricht2011}. Less simplified approximations have also been used for more complex geometries   \cite{Joyce2016,Bergren2014,Bergren2016,Parkinson2012,Nemec2010,Nemec2009a,Fekete2009}. All of these approximations are based on a number of assumptions and approximations in terms of the probe wavelength and the magnitude of the relative change in transmission/reflection due to photoexcitation. It is important to note that improper use of these simplified equations can give a significant deviation from the expected material parameters, which may in turn lead to incorrect conclusions regarding the underlying charge-carrier dynamics of the material \cite{Nienhuys2005,DAngelo2016,Hempel2017,Krewer2018}. However, assessing the appropriate approximation for a given sample is not always trivial, since there are no clear boundaries for validity. Despite this uncertainty, only a few groups have investigated the validity of these approximations in depth, and only for few specific cases in the reflection geometry \cite{DAngelo2016,Hempel2017}.
Using a 0.33 mm undoped GaAs wafer as a case study, D'Angelo et al. \cite{DAngelo2016} demonstrated that the approximations of a step-like excitation and a thin film approximation are generally insufficient for extracting the transient optical properties of thin films in the reflection geometry, due to the strong phase-sensitivity of these measurements. Similarly, Hempel et al. \cite{Hempel2017} investigated the effect of substrate refractive index and thickness on the sensitivity of thin film transient reflection measurements. By comparing thin films of polycrystalline Cu$_2$SnZnSe$_4$ placed on a glass substrate and on a highly conductive substrate of molybdenum, they showed that the thin film approximation breaks down when film and substrate refractive index differs significantly. While these studies highlight potential problems with specific approximations and sample geometries, it can be difficult to determine to what extent problems persist in other, more common sample geometries. 

In this paper, we present a general method for evaluating the accuracy of various approximations for any particular sample geometry. It should be noted that regimes of applicability can be evaluated using series expansions of the transmission functions for the given geometry, since many of the commonly used approximation rely on first order Taylor expansions. However, this is not straight forward even for the simple case of a single homogeneously photoexcited layer, see supplementary material section \ref{sec:Supp_taylor}. Instead, using a modified transfer matrix method, we simulate the transmitted and reflected fields of a photoexcited, multilayer structure, where the thickness of each layer and the photoinduced change in the complex permittivity are treated as input parameters. We then use these fields to extract approximate solutions for the complex permittivity of the material, which we compare to the input parameters, allowing us to assess ranges of validity. Using this method we calculate the relative deviation of the most commonly used approximations for a broad range of sample geometries, where the probe wavelength ranges from much smaller to much larger than the optical penetration depth of the pump beam in the material. Somewhat surprisingly, we find that these approximations are truly valid only in extreme cases where the optical thickness of the sample is several orders of magnitude smaller or larger than the probe wavelength. We then go beyond the most commonly applied approximations, using a general numerical approach based on the transfer matrix method, which exhibits a much broader range of validity. 

\section{The Transfer Matrix Method}
In a standard OPTP experiment the material of interest will typically be part of some multilayer structure. For example, the sample can be placed on a substrate or in a cuvette. On photoexcitation, part or all of this multilayer structure experiences a change in optical parameters, and the photoexcited change in the THz transmission/reflection of the entire structure is measured. Note that in this article we deal only with cases where the photoexcited changes to the optical parameters vary on a timescale which is slow compared to the period of the THz field, where the excited state optical parameters can be treated as quasi-static. In such circumstances, one wants a full wave solution capable of determining the complex transmission and reflection coefficients of arbitrary multilayers of differing index. The most straightforward approach here uses the \textit{transfer matrix method} (TMM). Below, we  summarize the important definitions and equations, but a full derivation can be found in many introductory textbooks on optics, for example Pedrotti et al. \cite[Chapter 22,]{Pedrotti2006}.

\begin{figure}
	\centering
	\includegraphics[width=1\linewidth]{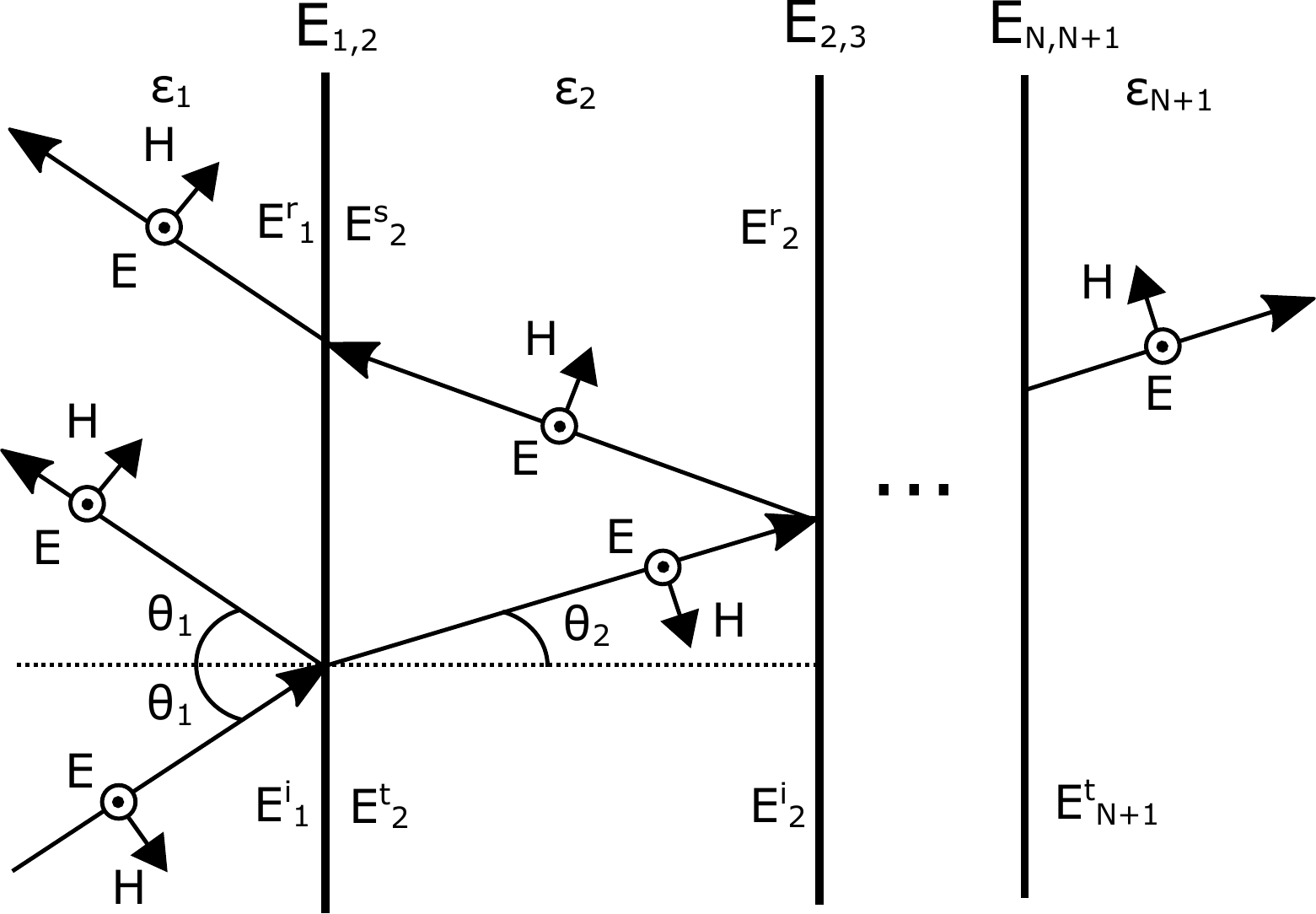}
	\caption{Wave propagation of TE-polarized field with incident angle $\theta_0$ through a multilayer structure of $N$ layers, each with a permittivity $\epsilon_j$, where $j=1,2,\dots,N+1$. The continuous THz field at each interface $E_{j,j+1}$ is determined by the incident ($E^i$, $E^s$), reflected ($E^r$) and transmitted ($E^t$) fields at each side of the interface, with $E^-_{j,j+1} = E^i_j+E^r_j$ and $E^+_{j,j+1} = E^t_{j+1}+E^s_{j+1}$ for the left and right side respectively.}
	\label{fig:tmm}
\end{figure}

Figure \ref{fig:tmm} shows a general $N$-layer structure consisting of homogeneous layers of thickness $d_j$ and complex permittivity $\epsilon_j$, where $j=1,2,3,\dots, N+1$. Here layer 1 and $N+1$ are the semi-infinite incident and transmitted regions, respectively. The electric ($E$) and magnetic ($H$) fields at interfaces $j$ and $j+1$ are related through
\begin{align}
\begin{pmatrix}
E_{j-1,j} \\
Z_0 H_{j-1,j}
\end{pmatrix} = 
\begin{pmatrix}
\cos{\delta_j} & \frac{-i \sin{\delta_j}}{\gamma_j} \\
-i\gamma_j\sin{\delta_j} & \cos{\delta_j}
\end{pmatrix}
\begin{pmatrix}
E_{j,j+1} \\ Z_0 H_{j,j+1}
\end{pmatrix},
\end{align}
where $Z_0$ is the free-space impedance, $\delta_j = \omega \sqrt{\epsilon_j} d_j \cos{\theta_j}/c$ is the complex phase (including absorption) accumulated through the $j$'th layer, $\theta_j$ is the incident angle on the $j$'th interface, $\omega$ is the angular frequency of the fields, $c=299792458$ m/s is the vacuum speed of light, and $\gamma^{TE}_j = \sqrt{\epsilon_j}\cos{\theta_j}$ and $\gamma^{TM}_j = \cos{\theta_j}/\sqrt{\epsilon_j}$ for TE- and TM-polarization, respectively. 
We define the transfer matrix for the $j$-th layer as
\begin{align}
M_j \equiv 
\begin{pmatrix}
\cos{\delta_j} & \frac{-i \sin{\delta_j}}{\gamma_j} \\
-i\gamma_j\sin{\delta_j} & \cos{\delta_j}
\end{pmatrix}.
\end{align}
Note that, in the absence of gain, since the off axis components of the transfer matrix have a negative sign, the formulation above is consistent with a positive sign for the imaginary part of the permittivity as per definition.	
The fields transmitted through the entire multilayer structure are related to the incident fields via the relation
\begin{align}
\begin{pmatrix}
E_{12} \\
Z_0 H_{12}
\end{pmatrix} &= 
M_1 M_2 \cdots M_N
\begin{pmatrix}
E_{N,N+1} \\
Z_0 H_{N,N+1}
\end{pmatrix} \nonumber\\
&= M_{tot}
\begin{pmatrix}
E_{N,N+1} \\
Z_0 H_{N,N+1}
\end{pmatrix}, \label{eq:Mtotal}
\end{align}
where $M_{tot}$ is the total transfer matrix of the system, given by
\begin{align}
M_{tot}=\begin{pmatrix}
m_{11} & m_{12} \\
m_{21} & m_{22}
\end{pmatrix}.
\end{align}
\eqref{eq:Mtotal} can be rewritten in terms of $E^i_1$, $E^r_1$ and $E^t_{N+1}$:
\begin{eqnarray}
\begin{pmatrix}
E^i_1+E^r_1 \\
(E^i_1-E^r_1)\gamma_1
\end{pmatrix} = M_{tot}
\begin{pmatrix}
E^t_{N+1} \\
\gamma_j E^t_{N+1}
\end{pmatrix}.\label{eq:TMM_total}
\end{eqnarray}
The reflection ($r$) and transmission ($t$) coefficients for the entire structure can then be found from \eqref{eq:TMM_total}
\begin{align}
r&=\frac{E^r_1}{E^i_1}=\frac{\gamma_1 m_{11}+\gamma_1\gamma_t m_{12}-m_{21}-\gamma_t m_{22}}{\gamma_1 m_{11} + \gamma_1\gamma_t m_{12} + m_{21} + \gamma_t m_{22}}, \label{eq:reflection}\\
t&=\frac{E^t_{N+1}}{E^i_1}=\frac{2\gamma_1}{\gamma_1 m_{11} + \gamma_1\gamma_t m_{12} + m_{21} + \gamma_t m_{22}}. \label{eq:transmission}
\end{align}
If $\epsilon_j$ and $d_j$ is known for each layer, it is fairly straight-forward to calculate the transmitted and reflected fields of the multilayer structure. The reverse is also possible, i.e. \eqref{eq:reflection} or \eqref{eq:transmission} can be solved for an unknown $\epsilon_j$, assuming the incident and transmitted/reflected fields are known, which is normally the case in OPTP experiments. However, without approximations to \eqref{eq:reflection} and \eqref{eq:transmission}   their solutions must be found numerically, which can be resource and time intensive, depending on the complexity of the multilayer structure.

\section{Wave-propagation in an optically pumped medium}
The goal of an OPTP experiment is normally to determine the photoinduced change in the THz optical properties of a sample. This is done by measuring either the transmitted or reflected electric field of the photoexcited sample $E_{exc}$, as well as an appropriate reference, usually the unexcited sample $E$. From these two measurements, the non-equilibrium permittivity of the photoexcited sample $\epsilon_{exc}$ can be determined by solving the equation
\begin{align}\label{eq:tn}
\frac{E_{exc}(\omega,\epsilon_{exc})}{E(\omega)} = \frac{t_{exc}(\omega,\epsilon_{exc})}{t(\omega)},
\end{align} 
where $t$ and $t_{exc}$ are the complex transmission functions of the unexcited and photoexcited multilayer structure. \eqref{eq:tn} is commonly written in terms of the change in transmitted electric field $\Delta{}E=E_{exc}-E$:
\begin{align}\label{eq:tn2}
\frac{\Delta{}E(\omega,\epsilon_{exc})}{E(\omega)}=\frac{t_{exc}(\omega,\epsilon_{exc})}{t(\omega)}-1.
\end{align} 

\begin{figure}
	\centering
	\includegraphics[width=0.8\linewidth]{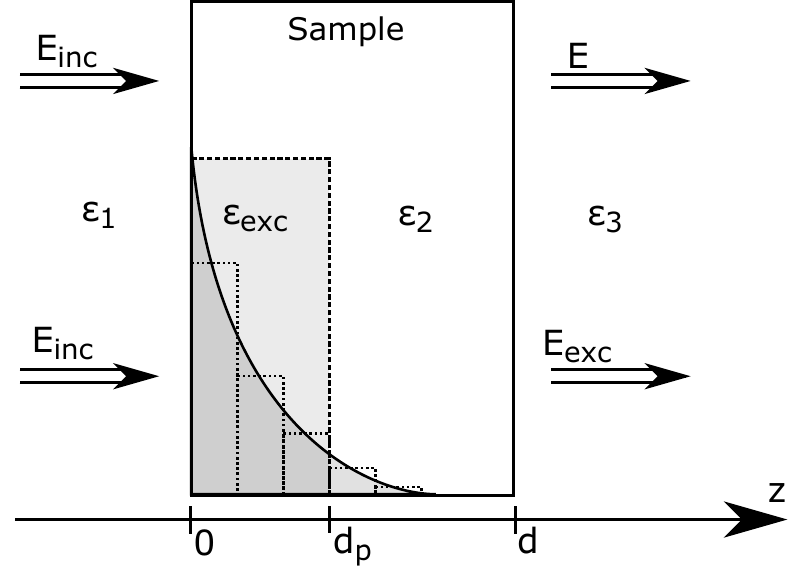}
	\caption{Diagram of typical pump-probe experimental scenario. A sample with permittivity $\epsilon_2$, surrounded by an incident ($\epsilon_1$) and transmitted medium ($\epsilon_3$), is photoexcited due to an incident pump pulse with penetration depth $d_p$, resulting in a new photoexcited permittivity at the surface of the sample $\epsilon_{exc}$. Here $E_{exc}$ and $E$ are the transmitted fields through the photoexcited and unexcited sample, respectively. The photoexcitation decays exponentially within the sample due to attenuation of the incident pump, where $d_p$ is the distance at which the pump intensity has decayed to $1/e$. We represent this by dividing the sample into $N$ homogeneous layers, each with a permittivity $\epsilon_{exc}(z)$ determined by the distance $z$ into the sample. This is commonly approximated as a single homogeneous layer of thickness $d_p$ and permittivity $\epsilon_{exc}$.}
	\label{fig:photoexc2}
\end{figure}

Assuming the equilibrium permittivity and thickness of each layer of the multilayer structure are known, and that the dynamics of relaxation are slow compared the period of the THz radiation, one can solve \eqref{eq:tn2} as a function of the non-equilibrium permittivity $\epsilon_{exc}$ for a given angular frequency $\omega$. In figure \ref{fig:photoexc2} we consider a common scenario of a sample with permittivity $\epsilon_2$, surrounded by an incidence ($\epsilon_1$) and transmission medium ($\epsilon_3$). If the sample is photoexcited, one expects a non-equilibrium permittivity that varies spatially through the sample, depending on the penetration of the pump pulse into the sample. 
If pump absorption is linear, and the amplitude of the THz permittivity also varies linearly with the density of photospecies, the photoinduced change in permittivity $\Delta\epsilon$ is expected to decay exponentially in the propagation direction, following the attenuation of the incident pump, and described by a decay length defined by a penetration depth $d_p$ of the pump light, as assumed in \cite{Kuzel2007,Kuzel2014a,Ulbricht2011}:
\begin{align}
\Delta\epsilon(z)&=\Delta\epsilon_{s} e^{-z/d_p} \label{eq:epsz},
\end{align}
where $\Delta\epsilon_{s}$ is the change in permittivity at the surface of the sample, which is assumed to be sufficiently small so that it scales linearly with the incident pump light. 
In this case, the spatially varying permittivity can be approximated by dividing the sample into $N$ homogeneous layers, each with a permittivity $\epsilon_{exc}(z)$ determined by the distance $z$ into the sample.
Using \eqref{eq:transmission} and \eqref{eq:tn2} and \eqref{eq:epsz} it is then possible to determine $\Delta{}\epsilon_s$ from the experimentally obtained $\Delta{}E/E$, given that thickness ($d$), penetration depth ($d_p$) and equilibrium permittivity ($\epsilon_2$) of the sample is known. However, no analytical solution exists for $\Delta{}\epsilon_s$ in its current form, so a solution must be obtained numerically, which can be resource and time consuming depending on the size of $N$.

A common approximation is to represent the photoexcitation as a single homogeneous layer of thickness $d_p$ and constant photoexcited permittivity $\epsilon_{exc}$, i.e. a step-like excitation approximation. In this case, depending on the size of the wavelength inside the sample $\lambda_s=\lambda_0/Re(\sqrt{\epsilon_2})$ compared to the thickness and penetration depth, various further approximations can be made, which result in a simpler and analytical solution for $\Delta{}\epsilon$ \cite{Kuzel2007,Nienhuys2005,Knoesel2004}. In order to obtain an analytical solution, the most common approximations assume a small perturbation of the photoexcited sample, i.e. $\Delta{}E/E<<1$, and that the wavelength is much shorter or much larger than the thickness and penetration depth of the sample, i.e. $|\beta|^{-1}=\frac{\lambda_0}{2\pi d |\sqrt{\epsilon_2}|}<<1$  \cite{Knoesel2004,Dakovski2007,Kunneman2013} or $\frac{\lambda_0}{2\pi d |\sqrt{\epsilon_2}|}>>1$  \cite{Nienhuys2005,Nienhuys2005a,Alberding2016,Jnawali2013a,Xing2017,Jin2014,Joyce2016,Nemec2015,Beaudoin2014,Lu2013,Tang2012,Yettapu2016,Jnawali2015,Pijpers2010,Xu2009}, resulting in
\begin{align}
\Delta{}\epsilon_{short}&=-\frac{2ic\sqrt{\epsilon_2}}{\omega d}\frac{\Delta{}E}{E}, &\text{(short $\lambda$ limit)} \label{eq:short}\\
\Delta{}\epsilon_{long}&=-\frac{ic(\sqrt{\epsilon_1}+\sqrt{\epsilon_3})}{\omega d}\frac{\Delta{}E}{E}, &\text{(long $\lambda$ limit)}. \label{eq:long}
\end{align}
\eqref{eq:short} is known as the short wavelength limit or thick sample approximation, and \eqref{eq:long} is known as the long wavelength limit or thin-film approximation. A full derivation can be found in appendix \ref{sec:A1}, along with equivalent expressions written in terms of either the complex refractive index or the complex conductivity. If Re$(\sqrt{\epsilon_2}) >$ Im$(\sqrt{\epsilon_2})$, then the conditions for the short and long wavelength limits approximately become $\lambda_s/d<<1$ and $\lambda_s/d>>1$, respectively. However, as we will show later on, $\lambda_s/d$ is actually not a good parameter for judging the validity of a given approximation and can be a bit misleading. Note that \eqref{eq:short} and \eqref{eq:long} assume a homogeneous excitation of the entire sample, which occurs for weakly absorbing materials, i.e. $d_p \approx d$. For strongly absorbing materials, i.e. $d_p << d$, similar expressions to \eqref{eq:short} \cite{Hendry2005,Hendry2006a,Ulbricht2011} and \eqref{eq:long} \cite{Cunningham2008,Cunningham2013,Strothkamper2012,Nemec2008,Lui2001,Kuzel2007,Jensen2014a,Terashige2015,Zhang2017,Petersen2017} can be derived, where $d$ is replaced by $d_p$. In this case, the photoexcited region is approximated as thin layer of thickness $d_p$, located at the surface of the sample.
\begin{figure*}
	\centering
	\includegraphics[width=1\linewidth]{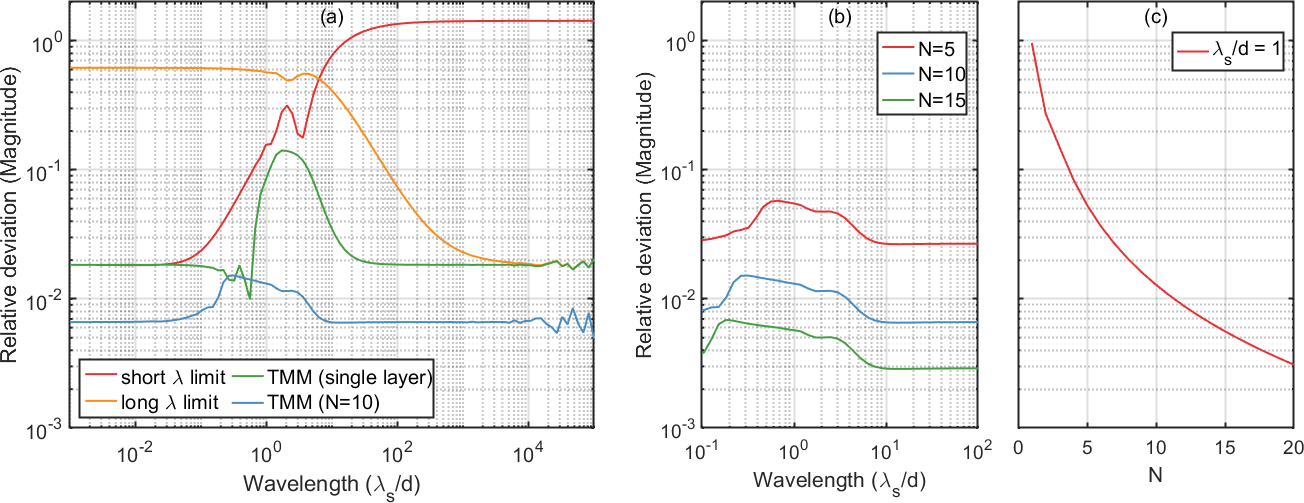}
	\caption{\textbf{(a)} Magnitude of the relative deviation $|(\Delta\epsilon_s^{out}-\Delta\epsilon_s^{in})/\Delta\epsilon_s^{in}|$ of the different approximations as a function of wavelength inside the sample $\lambda_s/d$. Here ``short $\lambda$ limit" and ``long $\lambda$ limit" are the short and long wavelength limit, ``TMM (N = 10)" is the numerical transfer matrix method approach with the photoexcited sample represented by $N = 10$ layers, and ``TMM single layer" uses the same transfer matrix method approach but with the photoexcitation being approximated as a single homogeneous layer of thickness $d_p$ and permittivity $\epsilon_{exc}$. Furthermore, the magnitude of the relative deviation of the transfer matrix method approach is plotted \textbf{(b)} as a function of $\lambda_s/d$ for $N=5,10,15$, and \textbf{(c)} as a function of $N$ for $\lambda_s/d=1$.}
	\label{fig:modeltest}
\end{figure*}
We now examine the relative deviation of the four different analysis approaches mentioned so far. Specifically: (i) the short wavelength limit of \eqref{eq:short} with $d=d_p$; (ii) the long wavelength limit of \eqref{eq:long} with $d=d_p$; (iii) the TMM approach with the photoexcited region of the sample approximated as a single homogeneous layer of thickness $d_p$ and permittivity $\epsilon_{exc}$; and (iv) the full TMM approach using $N=10$ layers to represent the photoexcited sample. In order to analyse the relative deviation of the different approximations, we compare results for a sample of known thickness $d$, penetration depth $d_p=d/4$, and equilibrium permittivity $\epsilon_2=10.7+6.8i$, with the incident region being air ($\epsilon_1=1$), and the transmitted region being a quartz substrate ($\epsilon_3=3.8$) \cite{Naftaly2007}. We assume a photoexcited sample response described by $\Delta{}\epsilon(z)$, and modelled according to \eqref{eq:epsz}, with $\Delta{}\epsilon_s=(1+1i)\times10^{-6}$. Note that such a sample is representative of a semiconductor thin-film deposited on a THz-transparent substrate. This type of sample has often been examined using terahertz spectroscopy, for example in the case of semiconductor nanowires \cite{Joyce2016}. As a simple example we have chosen the value of $\epsilon_2$ to be that of doped silicon \cite{Nashima2001} at 1 THz, since this semiconductor material has a noticeable absorption in the THz frequency range, and we have chosen $\Delta\epsilon_{s}$ to be sufficiently small so that $\Delta{}E/E<<1$ for all wavelengths ranging from $\lambda_s/d=10^{-3}$ to $\lambda_s/d=10^5$. We find that the transmission of the sample is essentially constant for $N>100$, hence we use calculations with $N=100$ layers to be representative of the true transmission. We then feed the generated $\Delta{}E/E$ into the various approximate analyses and compare the resulting $\Delta{}\epsilon_s^{out}$ with the input value of $\Delta{}\epsilon_s^{in}$. This gives a relative deviation $(\Delta\epsilon_s^{out}-\Delta\epsilon_s^{in})/\Delta\epsilon_s^{in}$ with which to assess the validity of the various approximations described above. Note that the code used for this analysis can be found on the ArXiv supplementary material repository for this article, and can be adapted for other sample geometries and tested using other approximations.

The resulting magnitude of the relative deviation is plotted in figure \ref{fig:modeltest}a as a function of the wavelength normalized to the thickness of the sample $\lambda_s/d$. Note that in this context, the magnitude of the relative deviation tells us how close $\Delta\epsilon_s^{out}$ is to $\Delta\epsilon_s^{in}$, while the phase tells us whether $\Delta\epsilon_s^{out}$ is smaller or larger than $\Delta\epsilon_s^{in}$ and if the ratio of the real and imaginary parts are distributed correctly. A corresponding phase plot of the data in figure \ref{fig:modeltest} can be seen in figure \ref{fig:modeltestarg} in the supplementary material. As one might expect, there is a significant error when the wavelength is comparable with the sample thickness, i.e. $\lambda_s/d\approx 1$, with up to 20--50\% in the magnitude of the relative deviation for the approximations in \eqref{eq:short} and \eqref{eq:long}, and 11\% relative deviation when using the TMM approach assuming a homogeneous photoexcited layer. The short and long wavelength approximations converge to a 2\% relative deviation for $\lambda_s/d<<1$ and $\lambda_s/d>>1$, respectively. This asymptotic value corresponds more or less exactly with the factor $1-(1-exp[-d/d_p])=exp[-d/d_p]=1.8\%$ found in eq. (27) from ref. \cite{Kuzel2007}, which the simple long- and short-wavelength limits do not account for. Indeed, when varying $d/d_p$, we observe that the asymptotic values shift accordingly, see supplementary material section \ref{sec:S_additional_plots} for more details. However, more surprisingly, the relative deviation of both the short and long wavelength approximations is considerably larger than we expected in the region $\lambda_s/d$ = $10^{-1}$ to $10^{3}$. For example, the relative deviation of the long wavelength limit, even when the wavelength is two orders larger than the sample thickness ($\lambda_s/d$ = $10^{2}$), is 10 \%. This underlines the problem in using these approximations without properly analysing their applicability for a given sample geometry, as has previously been observed by D'Angelo et al. \cite{DAngelo2016} and Hempel et al. \cite{Hempel2017}. Furthermore, we note that the error of these approximations will scale with the relative variation between $\epsilon_2$ and the surrounding mediums ($\epsilon_1, \epsilon_3$), in agreement with similar observations by Hempel et al. \cite{Hempel2017}, who showed that a high variation between sample and substrate refractive indices can lead to significant errors in the long wavelength approximation in the reflection geometry.
To avoid this issue, one can apply the TMM approach in full to better approximate the spatially varying profile of the non-equilibrium optical constants, as can be in figure \ref{fig:modeltest}a, where for $N=10$ we obtain a relative deviation of less than 2\% at $\lambda_s/d = 1$. In figures \ref{fig:modeltest}b-\ref{fig:modeltest}c we plot a similar analysis carried out using the TMM approach while varying the number of layers used to approximate the exponentially varying permittivity. We see that for $\lambda_s/d = 1$, the TMM approach converges relatively quickly for $N<5$ in terms of the magnitude of the relative deviation, but slows down beyond that, achieving a relative deviation less than of 1\% for $N\ge 12$. This demonstrates that the TMM approach is valid for all wavelengths as long as $N$ is large enough and thus TMM is a valid option in the case where no other approximation is applicable. However, we note that there is a trade off in terms of the computation time and memory requirement for this approach, which will scale non-linearly with $N$. Furthermore, when solving equations numerically there is always a chance that the obtained solution is not the correct one, but rather a local minimum, especially if the initial guess used in these routines is not close to the correct solution. To avoid this we solve \eqref{eq:tn2} numerically in terms of the photoexcited permittivity $\epsilon_{exc}$, using the unexcited permittivity $\epsilon_2$ as the initial guess, which is a fair approximation when $\Delta{}E/E<<1$ (and therefore $\Delta\epsilon_{s}/\epsilon_2<<1$). However in the far majority of cases, one of the approximations found in ref. \cite{Kuzel2007} can be used to extract the proper photoexcited permittivity of the sample, see supplementary material section \ref{sec:S_additional_plots} where we compare the performance of Eq. (27) from ref. \cite{Kuzel2007} with the long wavelength approximation. Unlike the TMM approach, the approximations found in ref. \cite{Kuzel2007} also account for phase matching and frequency mixing effects, which must be considered when the decay-time of the photoexcitation is similar to the duration of the THz pulse. However in this context, the method presented in this paper can be used as a quick comparison to work out which approximation is the most appropriate for a given sample geometry.

\begin{figure}
	\centering
	\includegraphics[width=1\linewidth]{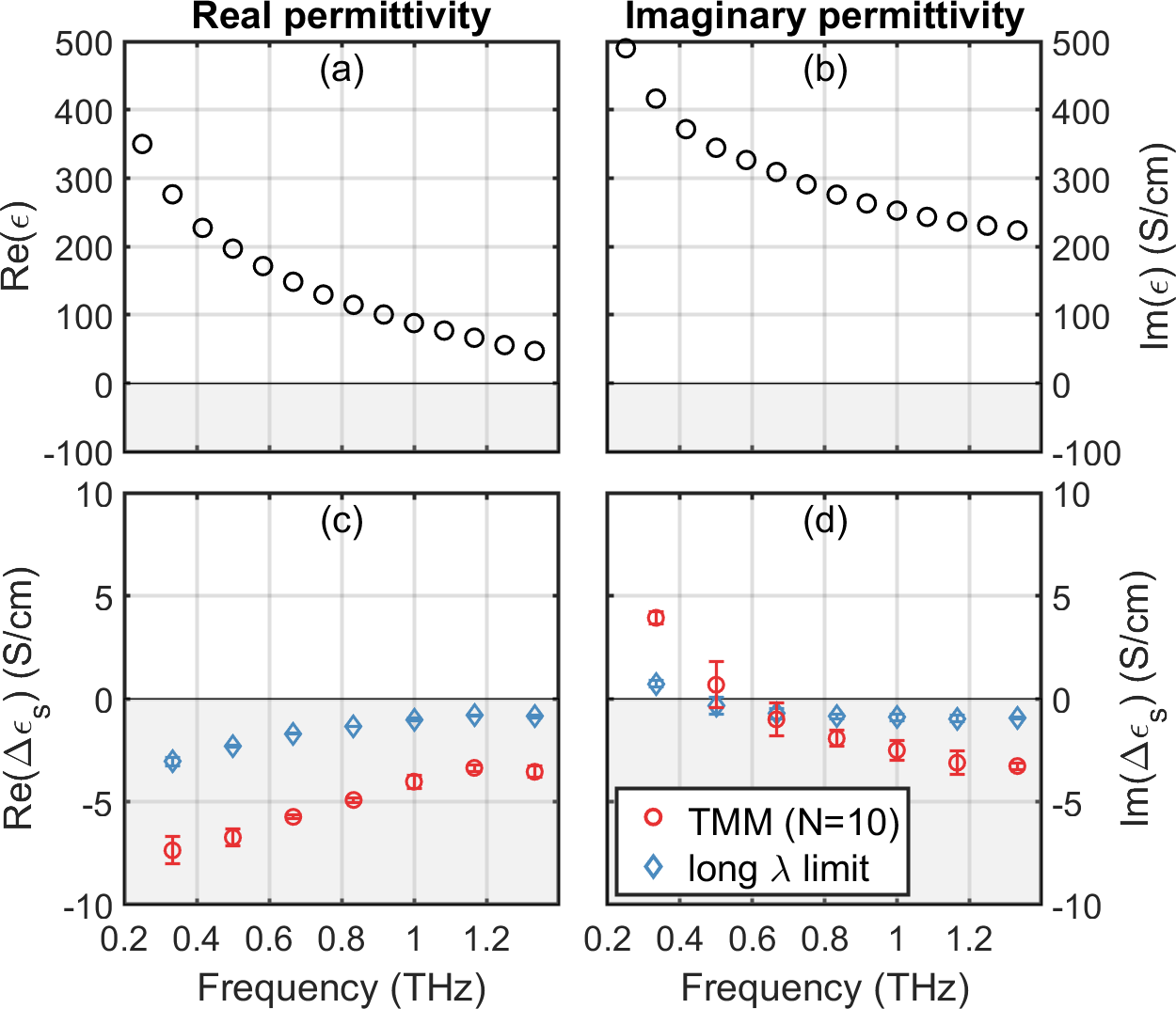}
	\caption{The unexcited permittivity \textbf{(a)} Re($\epsilon$) and \textbf{(b)} Im($\epsilon$) of a thin-film of carbon nanotubes placed on a quartz substrate. The film thickness is $d=700$ nm, penetration depth $d_p=285$ nm, corresponding to $\lambda_s/d = 37$ for 1 THz. The extracted photo-induced change in permittivity \textbf{(c)} Re($\Delta{}\epsilon_s$) and \textbf{(d)} Im($\Delta{}\epsilon_s$) due to 800 nm photoexcitation with a 100 fs pump pulse and incident fluence of 52 $\mu$J/cm$^2$, measured 10 ps after photoexcitation. We compare the resulting $\Delta{}\epsilon$ using the long wavelength limit with using the full TMM approach for $N=10$. The error-bars in \textbf{(c)}-\textbf{(d)} indicate the standard deviation of our data-points due to noise in our experimental setup.}
	\label{fig:experimentaldata}
\end{figure}
Finally in figure \ref{fig:experimentaldata} we examine a specific case of experimentally obtained data from a typical OPTP experiment \cite{Ulbricht2011}, in this case a film of single-walled carbon nanotubes of thickness 700 nm on a quartz substrate. We take $\Delta{}E/E$ measured two picoseconds after photoexcitation with an 800 nm, 100 fs pump pulse and incident fluence of 52 $\mu$J/cm$^2$, see \cite{Karlsen2018} for details of the sample preparation and the experimental setup. The sample geometry is the same as in figure \ref{fig:photoexc2}, where the permittivity of the sample at 1 THz is approximately $\epsilon_2=88+234i$, $\epsilon_1=1$ and $\epsilon_3=3.8$. In this case $\Delta{}E/E\le 0.5\%$, $\lambda_s/d=37$ for 1 THz, and the penetration depth $d_p$ is 285 nm. Observing the sample geometry (a thin film of thickness ten times thinner than the wavelength), one might assume the long wavelength limit in \eqref{eq:long} to be a valid approximation in this case. In figure \ref{fig:experimentaldata} we plot the obtained complex photo-induced change in permittivity $\Delta{}\epsilon_s$, extracted using the long wavelength limit, which we compare with the correct non-equilibrium permittivity $\Delta{}\epsilon_s$ obtained using the full TMM approach with $N = 10$, which we ensured had converged for this valued of $N$, see supplementary material section \ref{sec:Supp_convergence_tests} for details.  We see that the two methods give remarkably different spectral signals and magnitudes, illustrating the problems that can occur with improper use of approximate analysis methods. In this case, neither of the conditions $d_p/d<<1$ or $\lambda_0/(2\pi d \sqrt{\epsilon_2})>>1$ are properly satisfied for use of \eqref{eq:long}. Such a difference in the sample response could well result in misinterpretation of the photo-response, as discussed in ref. \cite{Nienhuys2005}. In this particular case, the change in sign of $\Delta{}\epsilon_s$ observed at 0.4 THz in figure \ref{fig:experimentaldata}b is indicative of a plasmon resonance in this material \cite{Karlsen2017,Karlsen2018,Slepyan2010a}, the parameters of which would be badly misinterpreted if one applied an approximate analysis.

\section{Conclusion}
In conclusion we have formulated a general method for evaluating the applicability of the most common approximation used in optical pump-THz probe spectroscopy. Somewhat surprisingly, we find that these approximations are truly valid only in extreme cases where the optical thickness of the sample is several orders of magnitude smaller or larger than the probe wavelength.This demonstrates the need to properly analyse the applicability of an approximation for a given sample geometry, since improper use of these approximations can lead to significant errors, which in turn can lead to an incorrect interpretation of the photoexcited response. 

\section*{Acknowledgements}
This research was partially supported by the European Union’s Seventh Framework Programme (FP7) for research, technological development and demonstration under project 607521 NOTEDEV. The Mathematica and Matlab code used for performing the Transfer Matrix calculations was initially adapted from the Mathematica code written by \href{http://sjbyrnes.com/multilayer-film-optics-programs/}{Steven Byrnes} \cite{Byrnes2016}. The Mathematica code used for this analysis can be found on the ArXiv supplementary material repository for this article.

\appendix
\renewcommand{\thefigure}{S\arabic{figure}}
\renewcommand{\thesection}{S\arabic{section}}
\setcounter{figure}{0}
\section{Derivation of the long- and short-wavelength limit}\label{sec:A1}
The short and long wavelength limits in \eqref{eq:short} and \eqref{eq:long} can be derived from the Fresnel equations under the assumption that the photo-induced change in the dielectric properties of the sample are small, i.e. $\Delta{}E/E<<1$. Furthermore, we assume a homogeneous excitation of the entire sample shown in figure \ref{fig:photoexc2}. 
When discussing the dielectric properties of a sample, it is common to consider either the complex refractive index, $n$, or the complex conductivity, $\sigma$, which are related to the permittivity, $\epsilon$, through
\begin{equation}\label{eq:cond}
n^2(\omega)= \epsilon(\omega) = 1 + \frac{i\sigma(\omega)}{\omega\epsilon_0},
\end{equation}
where $\epsilon_0=8.85\times10^{-12}Fm^{-1}$ is the vacuum permittivity and $\omega$ is the angular frequency. For the purpose of deriving \eqref{eq:short} and \eqref{eq:long}, we use the refractive index $n$, since it simplifies the equations.

The Fresnel transmission and reflection coefficients for normal incidence are given by
\begin{align}
t_{ij}&=\frac{2n_i}{n_i+n_j}, \\
r_{ij}&=\frac{n_i-n_j}{n_i+n_j}.
\end{align}
For the geometry shown in figure \ref{fig:photoexc2} the transmitted electric field through the photoexcited ($E_{exc}$) and unexcited ($E$) sample is then given by
\begin{align}
E&=\frac{t_{12}t_{23}e^{\beta_2}}{1-r_{21}r_{23}e^{2\beta_2}}, \label{eq:A1_unexc} \\
E_{exc}&=\frac{t_{1e}t_{e3}e^{\beta_e}}{1-r_{e1}r_{e3}e^{2\beta_e}}, \label{eq:A1_exc}
\end{align}
where the indices $e$ and $s$ denotes the photoexcited and unexcited sample, respectively, and $\beta_j=i\omega n_j d/c=i\delta_j$ is the accumulated phase through layer $j$. In \eqref{eq:A1_unexc} and \eqref{eq:A1_exc}, the numerator represents direct transmission through the sample, while the denominator represents multiple internal reflections in the sample. The relative change in the transmitted electric field $\Delta{}E/E$ is then given by
\begin{align}
\frac{\Delta{}E}{E}&=\frac{t_{1e}t_{e3}e^{\beta_e-\beta_2}}{t_{12}t_{23}}\frac{(1-r_{21}r_{23}e^{2\beta_2})}{(1-r_{e1}r_{e3}e^{2\beta_e})} - 1,\label{eq:deltaE} \\
&=\frac{t_{1e}t_{e3}e^{\beta_{\Delta}}}{t_{12}t_{23}}\frac{(1-r_{21}r_{23}e^{2\beta_2})}{(1-r_{e1}r_{e3}e^{2(\beta_2+\beta_{\Delta})})} - 1,\label{eq:deltaE1}
\end{align}
where $\beta_{\Delta}=i\omega\Delta{n}d/c$ and $\Delta{}n=n_{exc}-n_2$.

\eqref{eq:deltaE} cannot be solved analytically for $\Delta{}n$ in it's current form, however we can approximate the equation in several ways, depending on whether $\omega |n_2| d/c << 1$ (long wavelength limit) or $\omega |n_2| d/c >> 1$ (short wavelength limit). 

In case of the long wavelength limit, we can Taylor-expand all the exponentials in \eqref{eq:deltaE1}, since $|\beta_2|,|\beta_\Delta|<<1$:
\begin{align}\label{eq:deltaE2}
\frac{\Delta{}E}{E}=&\frac{t_{1e}t_{e3}(1+\beta_\Delta)}{t_{12}t_{23}} \nonumber\\
&\times\frac{1-r_{21}r_{23}(1+2\beta_2)}{1-r_{e1}r_{e3}(1+2\beta_2+2\beta_\Delta)} - 1
\end{align}
\eqref{eq:deltaE2} can then be reduced by discarding higher order terms of $\Delta{}n$, and subsequently solved for $\Delta{}n$: 
\begin{align}
\Delta{}n=\frac{c(c(n_1+n_3) + id(n_1-n_2)(n_2-n_3)\omega)}{d\omega(2icn_2+d(n_1-n_2)(n_3-n_2)\omega)}\frac{\Delta{}E}{E}\label{eq:long_derivation}
\end{align}

By substituting $x=\omega d/c$ in \eqref{eq:long_derivation} and Taylor-expanding in $x$ up to the first order, the equivalent approximation of \eqref{eq:long} is obtained in terms of $\Delta{}n$:
\begin{align}
\Delta{}n&\approx-\frac{n_1+n_3}{n_2}\frac{ic}{2\omega d}\frac{\Delta{}E}{E} &\text{(long $\lambda$ limit)} \label{eq:long_n}
\end{align} 
for $\omega d \rightarrow 0$.

In case of the short wavelength limit, the assumption $|\beta_2|<<1$ no longer holds, so a slighty different approach is required, which is outlined in ref. \cite{Knoesel2004} and repeated in this appendix. In order to simplify the derivation, we begin by assuming $n_3 = n_1$, i.e. the incident and transmitted medium are the same. It is possible to derive \eqref{eq:short} without this assumption, however the derivation becomes much more cumbersome while the final equation will be the same in either case. Using $n_3 = n_1$, \eqref{eq:deltaE1} becomes
\begin{align}
\frac{\Delta E}{E}&=\frac{t_{1e}t_{e1}e^{\beta_{\Delta}}}{t_{12}t_{21}}\frac{(1-(r_{21}e^{\beta_2})^2)}{(1-(r_{e1}e^{\beta_2}e^{\beta_{\Delta}})^2)} - 1,\label{eq:deltaE3}
\end{align}
To approximate \eqref{eq:deltaE3}, we must rely on the assumption $\Delta{}n/n<<1$. We begin by Taylor-expanding the exponential $e^{\beta_\Delta}$, since the assumption $\beta_\Delta<<1$ is still valid:
\begin{align}
\frac{\Delta E}{E}&=\frac{t_{1e}t_{e1}(1+\beta_\Delta)}{t_{12}t_{21}}\frac{(1-(r_{21}e^{\beta_2})^2)}{(1-(r_{e1}e^{\beta_2})^2(1+2\beta_{\Delta}))} - 1,\label{eq:deltaE4}
\end{align}
Similarly, the transmission and the multiple reflection terms can be approximated by Taylor expanding in terms of $\Delta{}n$: 
\begin{align}
\frac{t_{1e}t_{e1}}{t_{12}t_{21}} &\approx 1+\frac{n_1-n_2}{n_2(n_1+n_2)}\Delta{}n,\label{eq:transmissionterm}
\end{align}
and
\begin{align}
\frac{(1-(r_{21}e^{\beta_2})^2)}{(1-(r_{e1}e^{\beta_2})^2(1+2\beta_{\Delta}))} \approx 1-MR\times\Delta{}n  \label{eq:MR}
\end{align}
where the multiple reflection term $MR$ is given by
\begin{align}
MR=\frac{2ie^{2\beta_2}(n_2-n_1)(-2icn_1+d\omega(n_2^2-n_1^2))}{ce^{2\beta_2}(n_1-n_2)^2(n_1+n_2)-c(n_1+n_2)^3}
\end{align}
From equations \eqref{eq:deltaE4}, \eqref{eq:transmissionterm} and \eqref{eq:MR} we can then obtain an analytical solution for $\Delta{}n$:
\begin{align}
\Delta{}n = \left[\frac{i\omega d}{c}+\frac{n_2-n_1}{n_2(n_2+n_1)}+MR\right]^{-1}\frac{\Delta E}{E}, \label{eq:dn_Knoesel}
\end{align}
where the first term represents modification of the propagation through the sample, the second term represents changes in reflective losses at the two interfaces, and the third term represents the contribution from multiple reflections. For very thick samples, i.e. the short wavelength limit ($\omega d/c >> 1$), the propagation term dominates \eqref{eq:dn_Knoesel} and thus the other terms can be disgarded, giving us the equivalent short wavelength limit of \eqref{eq:short} written in terms of $n$: 
\begin{align}
\Delta{}n&\approx-\frac{ic}{\omega d}\frac{\Delta{}E}{E}  &\text{(short $\lambda$ limit)} \label{eq:short_n}
\end{align} 
for $\omega d \rightarrow \infty$.

\eqref{eq:short_n} and \eqref{eq:long_n} can be rewritten in terms of the complex permittivity using $\Delta\epsilon=(n_2+\Delta{}n)^2-n_2^2$, resulting in \eqref{eq:short} and \eqref{eq:long}:  
\begin{align}
\Delta{}\epsilon_{short}&=-\frac{2icn_2}{\omega d}\frac{\Delta{}E}{E}, \label{eq:short_eps}\\
\Delta{}\epsilon_{long}&=-\frac{ic(n_1+n_3)}{\omega d}\frac{\Delta{}E}{E}, \label{eq:long_eps}
\end{align}
where we have discarded the second order term of $\Delta{}E/E$ under the assumption that $\Delta{}E/E<<1$. Likewise, the complex conductivity can be found from \eqref{eq:short_eps} and \eqref{eq:long_eps} using $\Delta\sigma=-i\omega\epsilon_0\Delta\epsilon$:
\begin{align}
\Delta{}\sigma_{short}&=-\frac{2c\epsilon_0n_2}{d}\frac{\Delta{}E}{E}, \label{eq:short_cond}\\
\Delta{}\sigma_{long}&=-\frac{c\epsilon_0(n_1+n_3)}{d}\frac{\Delta{}E}{E}, \label{eq:long_cond}
\end{align} 
It is important to note that for the long wavelength limit in \eqref{eq:long_n}, \eqref{eq:long_eps} and \eqref{eq:long_cond} we have assumed that the wavelength is much larger than the thickness of the sample. However, it is straightforward to derive a similar expression, where we consider a thin photoexcited region of thickness $d<<\lambda$ at the surface of a semi-infinite sample, which is another common approximation in OPTP. In this case, $d$ becomes the thickness of the photoexcited region $d_p$. Likewise the $d$ in \eqref{eq:short_n}, \eqref{eq:short_eps} and \eqref{eq:short_cond} can be interchanged with the penetration depth $d_p$ in the case where we are no longer considering a homogeneous excitation of the entire sample.

\section{Assessment of Approximations through Taylor Expansion}\label{sec:Supp_taylor}
In this section we demonstrate an alternative method for assessing a given approximation by considering the higher order terms of the Taylor expansion of the transmission function for a given sample geometry. As an example we consider the simplest case of a homogeneously photoexcited sample in air, which is described by \eqref{eq:deltaE1}:
\begin{align}
\frac{E_{exc}}{E}&=\frac{t_{1e}t_{e1}e^{\beta_{\Delta}}}{t_{12}t_{21}}\frac{(1-(r_{21}e^{\beta_2})^2)}{(1-(r_{e1}e^{\beta_2}e^{\beta_{\Delta}})^2)}. \label{eq:transHom}
\end{align}
We established previously that for $|\beta_2|,|\beta_\Delta|<<1$, the long wavelength limit \eqref{eq:long} is a valid approximation. Since this approximation relies on Taylor expanding the exponentials in \eqref{eq:transHom} to the first order, we can assess the validity of this approximation by Taylor expanding \eqref{eq:transHom} in terms of $x=\omega d/c$:
\begin{align}
\frac{E_{exc}}{E}&=1+\frac{i(2n_2\Delta n + \Delta n^2)}{2n_1}x + O(x^2), \label{eq:transTaylor}
\end{align}
where we have refrained from writing the higher order terms explicitly here due to the complexity of these terms. Using the same input parameters as in the main manuscript (except $d_p>>d$ in this case), we then compare the different order terms in \eqref{eq:transTaylor} with the relative deviation of the long wavelength limit in figure \ref{fig:Taylor_test}:
\begin{figure}
	\centering
	\includegraphics[width=1\linewidth]{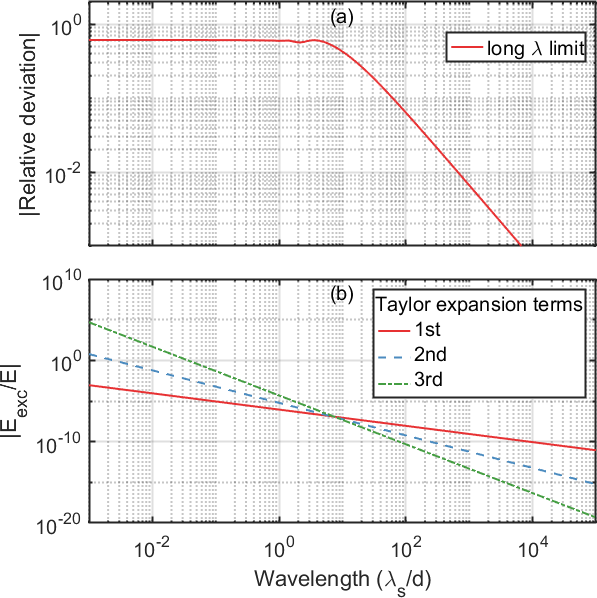}
	\caption{\textbf{(a)} Magnitude the relative deviation $(\Delta\epsilon_s^{out}-\Delta\epsilon_s^{in})/\Delta\epsilon_s^{in}$ as a function of wavelength for homogeneous excitation of the sample in air and extracted using the long wavelength limit. \textbf{(b)} The magnitude of the first, second and third order terms of the complex transmission function of the homogeneously excited sample, $E_{exc}/E$.}
	\label{fig:Taylor_test}
\end{figure}
We observe that the point at which the higher order terms start to dominate in figure \ref{fig:Taylor_test}b is more or less the same point at which the relative deviation of the long wavelength limit becomes massive. The benefit of this approach is that the dependence on the various sample parameters become much clearer, compared to the method presented in the main manuscript, however this method requires clear knowledge of the assumptions and derivation for each approximation, as well as writing out the Taylor expansion explicitly for transmission function of the sample geometry of interest, which can very quickly become tedious and massive for more complex geometries such as the one presented in figure \ref{fig:photoexc2}.

\section{Additional Approximations and Tests}\label{sec:S_additional_plots}
\begin{figure*}
	\centering
	\includegraphics[width=1\linewidth]{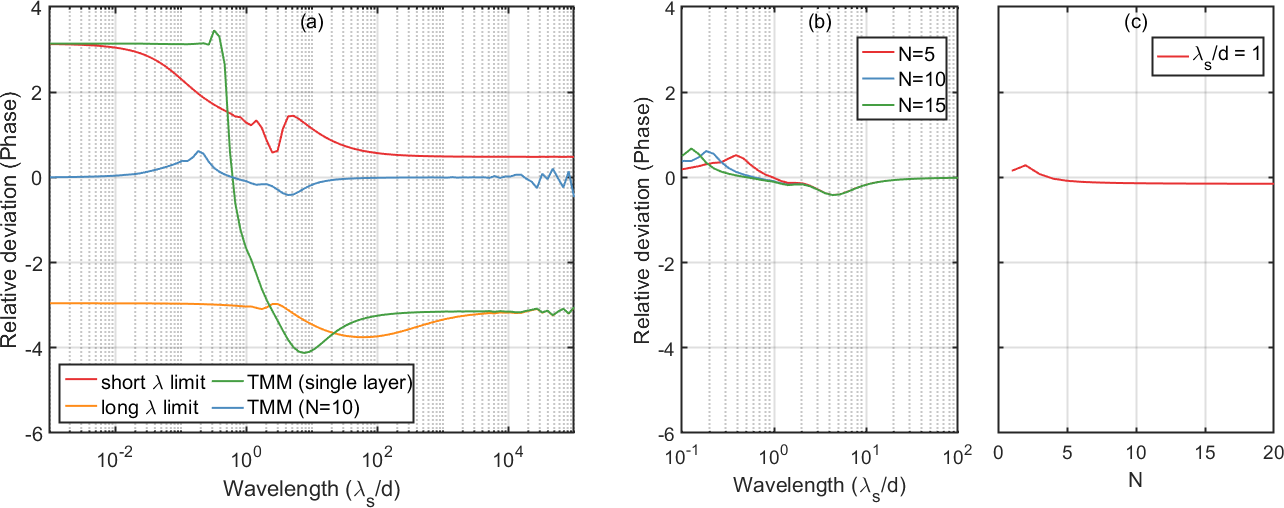}
	\caption{\textbf{(a)} Phase of the relative deviation $Arg((\Delta\epsilon_s^{out}-\Delta\epsilon_s^{in})/\Delta\epsilon_s^{in})$ of the different approximations as a function of wavelength inside the sample $\lambda_s/d$. Here ``short $\lambda$ limit" and ``long $\lambda$ limit" are the short and long wavelength limit, ``TMM (N = 10)" is the numerical transfer matrix method approach with the photoexcited sample represented by $N = 10$ layers, and ``TMM single layer" uses the same transfer matrix method approach but with the photoexcitation being approximated as a single homogeneous layer of thickness $d_p$ and permittivity $\epsilon_{exc}$. Furthermore, the phase of the relative deviation of the transfer matrix method approach is plotted \textbf{(b)} as a function of $\lambda_s/d$ for $N=5,10,15$, and \textbf{(c)} as a function of $N$ for $\lambda_s/d=1$.}
	\label{fig:modeltestarg}
\end{figure*}

In figure \ref{fig:modeltestarg}a we plot the phase of the relative deviation for the same data presented in figure \ref{fig:modeltest}, along with the convergence tests in figures \ref{fig:modeltestarg}b-\ref{fig:modeltestarg}c.
As mentioned in the main manuscript, it should be noted that in this context, the magnitude of the relative deviation tells us how close $\Delta\epsilon_s^{out}$ is to $\Delta\epsilon_s^{in}$, while the phase tells us whether $\Delta\epsilon_s^{out}$ is smaller or larger than $\Delta\epsilon_s^{in}$ and if the ratio of the real and imaginary parts are distributed correctly. For example, if $\Delta\epsilon_s^{out} < \Delta\epsilon_s^{in}$, but the ratio of their real and imaginary parts are the same, then the phase of their relative deviation will be $\pm\pi$. 

To verify the asymptotic behaviour of the short and long wavelength limits in figure \ref{fig:modeltest}, we compare the same approximations for different values of $d/d_p$ in figure \ref{fig:dp_test}:
\begin{figure}
	\centering
	\includegraphics[width=1\linewidth]{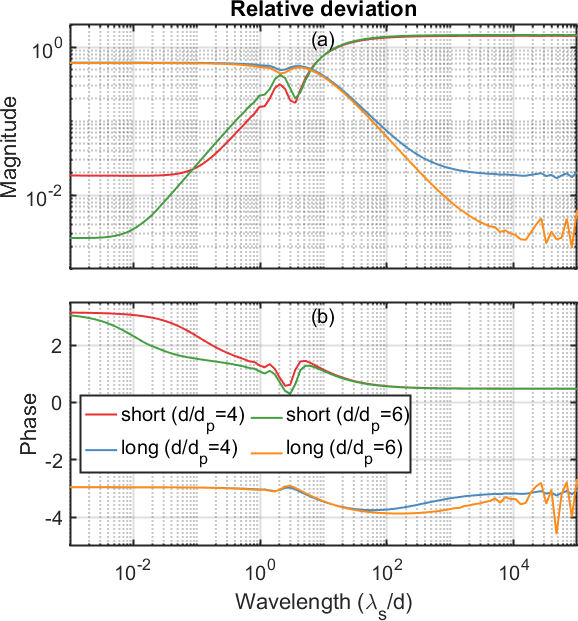}
	\caption{\textbf{(a)} Magnitude and \textbf{(b)} phase of the relative deviation $(\Delta\epsilon_s^{out}-\Delta\epsilon_s^{in})/\Delta\epsilon_s^{in}$ as a function of wavelength for different input values of $d/d_p$ and extracted using the short and long wavelength limits. The asymptotic value of the approximations are 1.8\% and 0.27\% for $d/d_p = 4$ and $d/d_p = 6$, respectively, which fits more or less exactly with the factor exp($-d/d_p$).}
	\label{fig:dp_test}
\end{figure}
We observe that the asymptotic value of the two approximations fit more or less exactly with the factor exp($-d/d_p$) for the different values of $d/d_p$, which is present in Eq. (27) and similar equations from ref. \cite{Kuzel2007} that accounts for when $d\approx d_p$. This indicates that the approximations presented in \cite{Kuzel2007} are generally more accurate widely applicable, which we demonstrate in figure \ref{fig:Kuzel_compare}:
\begin{figure}
	\centering
	\includegraphics[width=1\linewidth]{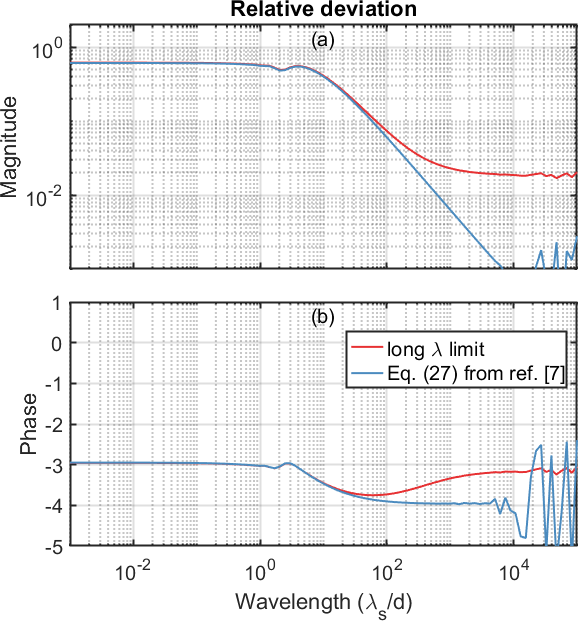}
	\caption{\textbf{(a)} Magnitude and \textbf{(b)} phase of the relative deviation $(\Delta\epsilon_s^{out}-\Delta\epsilon_s^{in})/\Delta\epsilon_s^{in}$ as a function of wavelength for $d/d_p = 4$, extracted using the long wavelength limit and Eq. (27) from ref. \cite{Kuzel2007}.}
	\label{fig:Kuzel_compare}
\end{figure}
Here Eq. (27) from ref. \cite{Kuzel2007} can be written in terms of $\Delta\epsilon$ as 
\begin{align}
\Delta\epsilon_{Eq. 27} = \frac{ic(n_1+n_3)}{\omega d_p (1-e^{-d/d_p})}\frac{\Delta E}{E}, \label{eq:Kuzel_eq27}
\end{align}
where we have ignored frequency-mixing and phase-matching effects. Interestingly, while \eqref{eq:Kuzel_eq27} is more accurate for large $\lambda_s/d>10^2$, below this value the two approximations give the same result, indicating that there is still a large wavelength range for which the more complex \eqref{eq:Kuzel_eq27} also becomes unreliable.

\section{Convergence Tests}\label{sec:Supp_convergence_tests}
\begin{figure}
	\centering
	\includegraphics[width=1\linewidth]{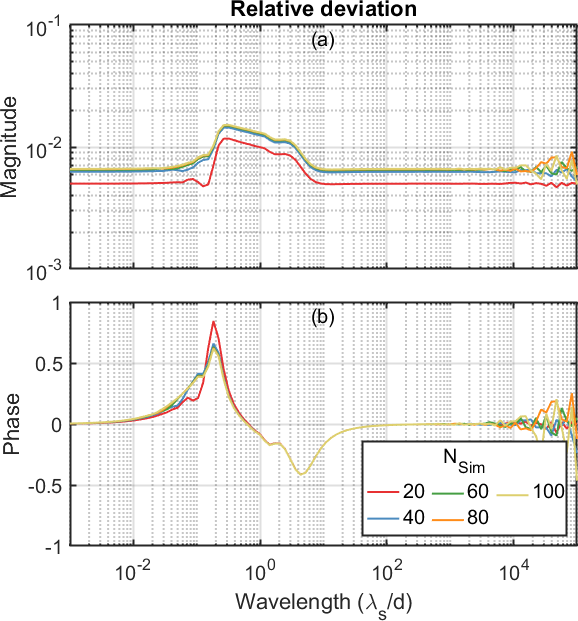}
	\caption{\textbf{(a)} Magnitude and \textbf{(b)} phase of the relative deviation $(\Delta\epsilon_s^{out}-\Delta\epsilon_s^{in})/\Delta\epsilon_s^{in}$ as a function of wavelength inside the sample $\lambda_s/d$, obtained using the TMM approach for $N=10$ layers. Here $N_{Sim}$ indicates the number of layers used to model the photoexcitation through the sample when the transmitted fields were initially simulated.}
	\label{fig:NSim_convergence_test}
\end{figure}

To ensure the validity of our TMM simulation, we tested how our results converge as we increase the number of layers $N_{sim}$ used when initially simulating the transmitted fields, see figure \ref{fig:NSim_convergence_test}. We observe that the results fairly quickly converge for $N_{sim}$ greater than approximately 40-60 layers, and become more or less indistinguishable when  $N_{sim}$ approaches 100 layers. 

To check the validity of our experimental data, we performed a similar convergence test with respect to the number of layers $N$ used in our TMM approach to extract our experimental data in figure \ref{fig:experimentaldata}:
\begin{figure}
	\centering
	\includegraphics[width=1\linewidth]{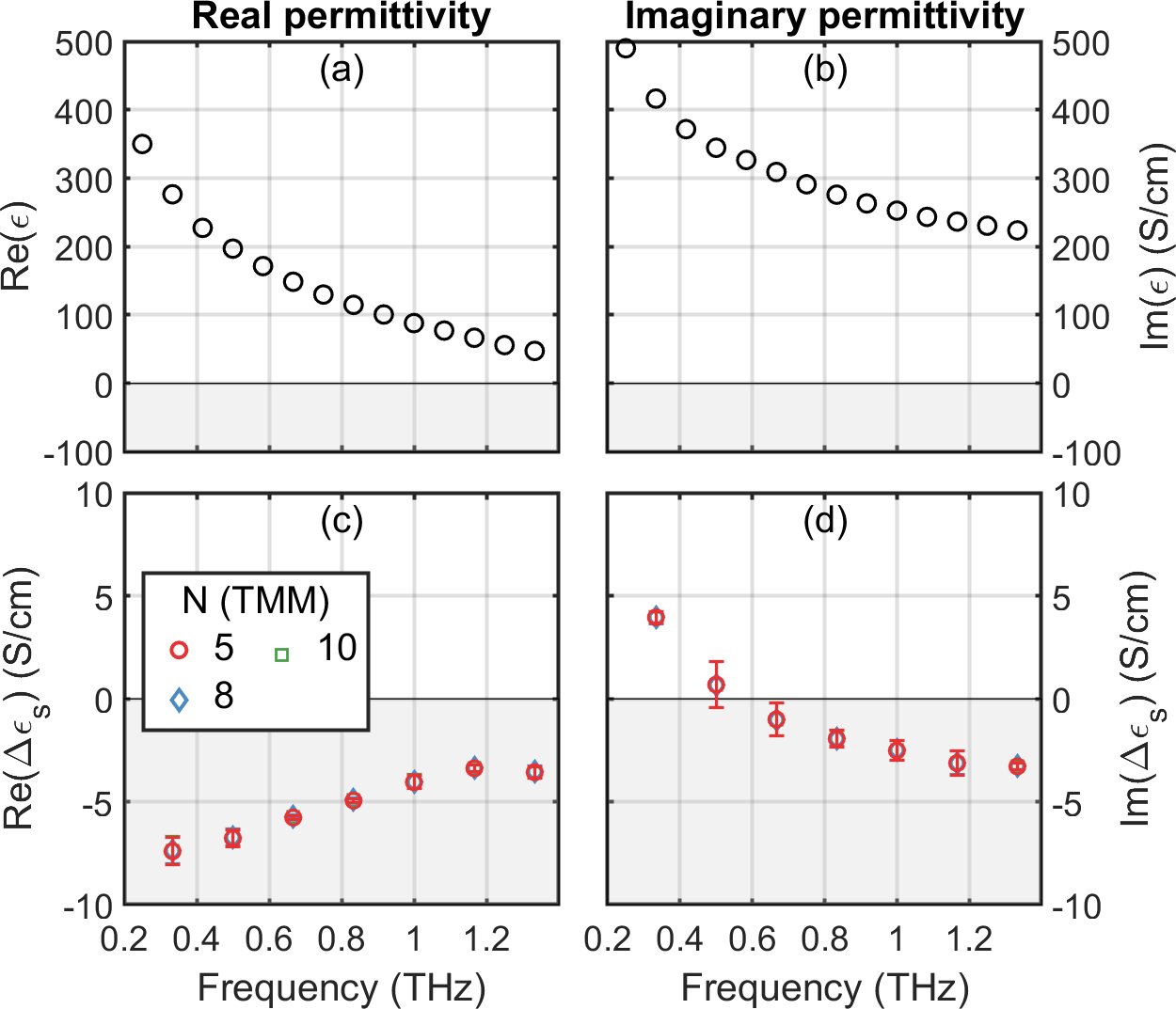}
	\caption{The unexcited permittivity \textbf{(a)} Re($\epsilon$) and \textbf{(b)} Im($\epsilon$) of a thin-film of carbon nanotubes placed on a quartz substrate. The film thickness is $d=700$ nm, penetration depth $d_p=285$ nm, corresponding to $\lambda_s/d = 37$ for 1 THz. The extracted photo-induced change in permittivity \textbf{(c)} Re($\Delta{}\epsilon_s$) and \textbf{(d)} Im($\Delta{}\epsilon_s$) due to 800 nm photoexcitation with a 100 fs pump pulse and incident fluence of 52 $\mu$J/cm$^2$, measured 10 ps after photoexcitation. We test the convergence of TMM approach for $N\in[5,10]$ and observe that the solutions are practically indistinguishable for these values of $N$. The error-bars in \textbf{(c)}-\textbf{(d)} indicate the standard deviation of our data-points due to noise in our experimental setup.}
	\label{fig:experimentaldata_Ntest}
\end{figure}
As can be seen in figure \ref{fig:experimentaldata_Ntest}, the data points for $N\ge5$ are practically indistinguishable, which fits with our expected relative deviation of a few percent seen in figure \ref{fig:modeltest}b. This means that the TMM solution shown in figure \ref{fig:experimentaldata} is more or less "exact" in this context.  

\clearpage
\bibliography{literature}

\begin{thebibliography}{10}
\newcommand{\enquote}[1]{``#1''}

\bibitem{Ulbricht2011}
R.~Ulbricht, E.~Hendry, J.~Shan, T.~F. Heinz, and M.~Bonn, \enquote{{Carrier
  dynamics in semiconductors studied with time-resolved terahertz
  spectroscopy},} {\protect\JournalTitle{Reviews of Modern Physics}}
  \textbf{83}, 543--586 (2011).

\bibitem{Jepsen2011}
P.~Jepsen, D.~Cooke, and M.~Koch, \enquote{{Terahertz spectroscopy and imaging
  - Modern techniques and applications},} {\protect\JournalTitle{Laser {\&}
  Photonics Reviews}} \textbf{5}, 124--166 (2011).

\bibitem{Joyce2016}
H.~J. Joyce, J.~L. Boland, C.~L. Davies, S.~A. Baig, and M.~B. Johnston,
  \enquote{{Electrical properties of semiconductor nanowires: Insights gained
  from terahertz conductivity spectroscopy.}}
  {\protect\JournalTitle{Semiconductor Science and Technology}} \textbf{31},
  1--21 (2016).

\bibitem{Schmuttenmaer2004}
C.~A. Schmuttenmaer, \enquote{{Exploring dynamics in the far-infrared with
  terahertz spectroscopy},} {\protect\JournalTitle{Chemical Reviews}}
  \textbf{104}, 1759--1779 (2004).

\bibitem{Duvillaret1996a}
L.~Duvillaret, F.~Garet, and J.-L.~L. Coutaz, \enquote{{A reliable method for
  extraction of material parameters in terahertz time-domain spectroscopy},}
  {\protect\JournalTitle{IEEE Journal of Selected Topics in Quantum
  Electronics}} \textbf{2}, 739--746 (1996).

\bibitem{Nienhuys2005}
H.~K. Nienhuys and V.~Sundstr{\"{o}}m, \enquote{{Intrinsic complications in the
  analysis of optical-pump, terahertz probe experiments},}
  {\protect\JournalTitle{Physical Review B - Condensed Matter and Materials
  Physics}} \textbf{71}, 1--8 (2005).

\bibitem{Kuzel2007}
P.~Ku{\v{z}}el, F.~Kadlec, and H.~N{\v{e}}mec, \enquote{{Propagation of
  terahertz pulses in photoexcited media: Analytical theory for layered
  systems},} {\protect\JournalTitle{The Journal of Chemical Physics}}
  \textbf{127}, 024506 (2007).

\bibitem{Bergren2016}
M.~R. Bergren, P.~K.~B. Palomaki, N.~R. Neale, T.~E. Furtak, and M.~C. Beard,
  \enquote{{Size-Dependent Exciton Formation Dynamics in Colloidal Silicon
  Quantum Dots},} {\protect\JournalTitle{ACS Nano}} \textbf{10}, 2316--2323
  (2016).

\bibitem{Huber2005}
R.~Huber, C.~K{\"{u}}bler, S.~T{\"{u}}bel, A.~Leitenstorfer, Q.~T. Vu, H.~Haug,
  F.~K{\"{o}}hler, and M.-C. Amann, \enquote{{Femtosecond Formation of Coupled
  Phonon-Plasmon Modes in InP: Ultrabroadband THz Experiment and Quantum
  Kinetic Theory},} {\protect\JournalTitle{Physical Review Letters}}
  \textbf{94}, 027401 (2005).

\bibitem{Jensen2014a}
S.~A. Jensen, K.-J. Tielrooij, E.~Hendry, M.~Bonn, I.~Rychetsk{\'{y}}, and
  H.~N{\v{e}}mec, \enquote{{Terahertz Depolarization Effects in Colloidal
  TiO$_2$ Films Reveal Particle Morphology},} {\protect\JournalTitle{The
  Journal of Physical Chemistry C}} \textbf{118}, 1191--1197 (2014).

\bibitem{Beard2000}
M.~C. Beard, G.~M. Turner, and C.~A. Schmuttenmaer, \enquote{{Transient
  photoconductivity in GaAs as measured by time-resolved terahertz
  spectroscopy},} {\protect\JournalTitle{Physical Review B - Condensed Matter
  and Materials Physics}} \textbf{62}, 15764--15777 (2000).

\bibitem{Hempel2017}
H.~Hempel, T.~Unold, and R.~Eichberger, \enquote{{Measurement of charge carrier
  mobilities in thin films on metal substrates by reflection time resolved
  terahertz spectroscopy},} {\protect\JournalTitle{Optics Express}}
  \textbf{25}, 17227 (2017).

\bibitem{Hempel2016}
H.~Hempel, A.~Redinger, I.~Repins, C.~Moisan, G.~Larramona, G.~Dennler,
  M.~Handwerg, S.~F. Fischer, R.~Eichberger, and T.~Unold, \enquote{{Intragrain
  charge transport in kesterite thin films—Limits arising from carrier
  localization},} {\protect\JournalTitle{Journal of Applied Physics}}
  \textbf{120}, 175302 (2016).

\bibitem{Cunningham2008}
P.~D. Cunningham and L.~M. Hayden, \enquote{{Carrier Dynamics Resulting from
  Above and Below Gap Excitation of P3HT and P3HT/PCBM Investigated by
  Optical-Pump Terahertz-Probe Spectroscopy},} {\protect\JournalTitle{The
  Journal of Physical Chemistry C}} \textbf{112}, 7928--7935 (2008).

\bibitem{Zajac2014}
V.~Zajac, H.~N{\v{e}}mec, C.~Kadlec, K.~Kůsov{\'{a}}, I.~Pelant, and
  P.~Ku{\v{z}}el, \enquote{{THz photoconductivity in light-emitting
  surface-oxidized Si nanocrystals: the role of large particles},}
  {\protect\JournalTitle{New Journal of Physics}} \textbf{16}, 093013 (2014).

\bibitem{Xiao2015}
Y.~Xiao, Z.-H. Zhai, Q.-W. Shi, L.-G. Zhu, J.~Li, W.-X. Huang, F.~Yue, Y.-Y.
  Hu, Q.-X. Peng, and Z.-R. Li, \enquote{{Ultrafast terahertz modulation
  characteristic of tungsten doped vanadium dioxide nanogranular film revealed
  by time-resolved terahertz spectroscopy},} {\protect\JournalTitle{Applied
  Physics Letters}} \textbf{107}, 031906 (2015).

\bibitem{Ziwritsch2016}
M.~Ziwritsch, S.~M{\"{u}}ller, H.~Hempel, T.~Unold, F.~F. Abdi, R.~van~de Krol,
  D.~Friedrich, and R.~Eichberger, \enquote{{Direct Time-Resolved Observation
  of Carrier Trapping and Polaron Conductivity in BiVO$_4$},}
  {\protect\JournalTitle{ACS Energy Letters}} \textbf{1}, 888--894 (2016).

\bibitem{Strothkamper2012}
C.~Strothk{\"{a}}mper, K.~Schwarzburg, R.~Sch{\"{u}}tz, R.~Eichberger, and
  A.~Bartelt, \enquote{{Multiple-Trapping Governed Electron Transport and
  Charge Separation in ZnO/In 2 S 3 Core/Shell Nanorod Heterojunctions},}
  {\protect\JournalTitle{The Journal of Physical Chemistry C}} \textbf{116},
  1165--1173 (2012).

\bibitem{Nemec2009}
H.~N{\v{e}}mec, H.-K. Nienhuys, E.~Perzon, F.~Zhang, O.~Ingan{\"{a}}s,
  P.~Ku{\v{z}}el, and V.~Sundstr{\"{o}}m, \enquote{{Ultrafast conductivity in a
  low-band-gap polyphenylene and fullerene blend studied by terahertz
  spectroscopy},} {\protect\JournalTitle{Physical Review B}} \textbf{79},
  245326 (2009).

\bibitem{Nemec2008}
H.~N{\v{e}}mec, L.~Fekete, F.~Kadlec, P.~Ku{\v{z}}el, M.~Martin, J.~Mangeney,
  J.~C. Delagnes, and P.~Mounaix, \enquote{{Ultrafast carrier dynamics in
  ${Br^+}$-bombarded InP studied by time-resolved terahertz spectroscopy},}
  {\protect\JournalTitle{Physical Review B}} \textbf{78}, 235206 (2008).

\bibitem{Strothkamper2013}
C.~Strothk{\"{a}}mper, A.~Bartelt, P.~Sippel, T.~Hannappel, R.~Sch{\"{u}}tz,
  and R.~Eichberger, \enquote{{Delayed Electron Transfer through Interface
  States in Hybrid ZnO/Organic-Dye Nanostructures},} {\protect\JournalTitle{The
  Journal of Physical Chemistry C}} \textbf{117}, 17901--17908 (2013).

\bibitem{Nemec2013}
H.~N\v{e}mec, V.~Zajac, I.~Rychetsky, D.~Fattakhova-Rohlfing, B.~Mandlmeier,
  T.~Bein, Z.~Mics, and P.~Kuzel, \enquote{{Charge Transport in {$TiO_2$} Films
  With Complex Percolation Pathways Investigated by Time-Resolved Terahertz
  Spectroscopy},} {\protect\JournalTitle{IEEE Transactions on Terahertz Science
  and Technology}} \textbf{3}, 302--313 (2013).

\bibitem{Liu2012}
H.~W. Liu, L.~M. Wong, S.~J. Wang, S.~H. Tang, and X.~H. Zhang,
  \enquote{{Ultrafast insulator–metal phase transition in vanadium dioxide
  studied using optical pump–terahertz probe spectroscopy},}
  {\protect\JournalTitle{Journal of Physics: Condensed Matter}} \textbf{24},
  415604 (2012).

\bibitem{HynekNemec2008}
H.~N{\v{e}}mec, H.~K. Nienhuys, F.~Zhang, O.~Inganas, A.~Yartsev, and
  V.~Sundstr{\"{o}}m, \enquote{{Charge carrier dynamics in alternating
  polyfluorene copolymer: Fullerene blends probed by terahertz spectroscopy},}
  {\protect\JournalTitle{Journal of Physical Chemistry C}} \textbf{112},
  6558--6563 (2008).

\bibitem{Cunningham2013}
P.~D. Cunningham, \enquote{{Accessing Terahertz Complex Conductivity Dynamics
  in the Time-Domain},} {\protect\JournalTitle{IEEE Transactions on Terahertz
  Science and Technology}} \textbf{3}, 494--498 (2013).

\bibitem{Jnawali2013a}
G.~Jnawali, Y.~Rao, H.~Yan, and T.~F. Heinz, \enquote{{Observation of a
  transient decrease in terahertz conductivity of single-layer graphene induced
  by ultrafast optical excitation.}} {\protect\JournalTitle{Nano letters}}
  \textbf{13}, 524--30 (2013).

\bibitem{Terashige2015}
T.~Terashige, H.~Yada, Y.~Matsui, T.~Miyamoto, N.~Kida, and H.~Okamoto,
  \enquote{{Temperature and carrier-density dependence of electron-hole
  scattering in silicon investigated by optical-pump terahertz-probe
  spectroscopy},} {\protect\JournalTitle{Physical Review B}} \textbf{91},
  241201 (2015).

\bibitem{Yettapu2016}
G.~R. Yettapu, D.~Talukdar, S.~Sarkar, A.~Swarnkar, A.~Nag, P.~Ghosh, and
  P.~Mandal, \enquote{{Terahertz Conductivity within Colloidal CsPbBr$_3$
  Perovskite Nanocrystals: Remarkably High Carrier Mobilities and Large
  Diffusion Lengths},} {\protect\JournalTitle{Nano Letters}} \textbf{16},
  4838--4848 (2016).

\bibitem{Lui2001}
K.~P.~H. Lui and F.~A. Hegmann, \enquote{{Ultrafast carrier relaxation in
  radiation-damaged silicon on sapphire studied by
  optical-pump–terahertz-probe experiments},} {\protect\JournalTitle{Applied
  Physics Letters}} \textbf{78}, 3478--3480 (2001).

\bibitem{Prasankumar2005}
R.~P. Prasankumar, A.~Scopatz, D.~J. Hilton, A.~J. Taylor, R.~D. Averitt, J.~M.
  Zide, and A.~C. Gossard, \enquote{{Carrier dynamics in self-assembled ErAs
  nanoislands embedded in GaAs measured by optical-pump terahertz-probe
  spectroscopy},} {\protect\JournalTitle{Applied Physics Letters}} \textbf{86},
  201107 (2005).

\bibitem{Minami2015}
Y.~Minami, K.~Horiuchi, K.~Masuda, J.~Takeda, and I.~Katayama,
  \enquote{{Terahertz dielectric response of photoexcited carriers in Si
  revealed via single-shot optical-pump and terahertz-probe spectroscopy},}
  {\protect\JournalTitle{Applied Physics Letters}} \textbf{107}, 171104 (2015).

\bibitem{Cooke2012}
D.~G. Cooke, A.~Meldrum, and P.~{Uhd Jepsen}, \enquote{{Ultrabroadband
  terahertz conductivity of Si nanocrystal films},}
  {\protect\JournalTitle{Applied Physics Letters}} \textbf{101}, 211107 (2012).

\bibitem{Cooke2012a}
D.~G. Cooke, F.~C. Krebs, and P.~U. Jepsen, \enquote{{Direct Observation of
  Sub-100 fs Mobile Charge Generation in a Polymer-Fullerene Film},}
  {\protect\JournalTitle{Physical Review Letters}} \textbf{108}, 056603 (2012).

\bibitem{Valverde-Chavez2015}
D.~A. Valverde-Ch{\'{a}}vez, C.~S. Ponseca, C.~C. Stoumpos, A.~Yartsev, M.~G.
  Kanatzidis, V.~Sundstr{\"{o}}m, and D.~G. Cooke, \enquote{{Intrinsic
  femtosecond charge generation dynamics in single crystal
  CH$_3$NH$_3$PbI$_3$},} {\protect\JournalTitle{Energy {\&} Environmental
  Science}} \textbf{8}, 3700--3707 (2015).

\bibitem{Nienhuys2005a}
H.-K. Nienhuys and V.~Sundstr{\"{o}}m, \enquote{{Influence of plasmons on
  terahertz conductivity measurements},} {\protect\JournalTitle{Applied Physics
  Letters}} \textbf{87}, 012101 (2005).

\bibitem{Alberding2016}
B.~G. Alberding, A.~J. Biacchi, A.~R. {Hight Walker}, and E.~J. Heilweil,
  \enquote{{Charge Carrier Dynamics and Mobility Determined by Time-Resolved
  Terahertz Spectroscopy on Films of Nano-to-Micrometer-Sized Colloidal Tin(II)
  Monosulfide},} {\protect\JournalTitle{The Journal of Physical Chemistry C}}
  \textbf{120}, 15395--15406 (2016).

\bibitem{Xing2017}
X.~Xing, L.~Zhao, Z.~Zhang, X.~Liu, K.~Zhang, Y.~Yu, X.~Lin, H.~Y. Chen, J.~Q.
  Chen, Z.~Jin, J.~Xu, and G.-h. Ma, \enquote{{Role of Photoinduced Exciton in
  the Transient Terahertz Conductivity of Few-Layer WS$_2$ Laminate},}
  {\protect\JournalTitle{The Journal of Physical Chemistry C}} \textbf{121},
  20451--20457 (2017).

\bibitem{Jin2014}
Z.~Jin, D.~Gehrig, C.~Dyer-Smith, E.~J. Heilweil, F.~Laquai, M.~Bonn, and
  D.~Turchinovich, \enquote{{Ultrafast Terahertz Photoconductivity of
  Photovoltaic Polymer–Fullerene Blends: A Comparative Study Correlated with
  Photovoltaic Device Performance},} {\protect\JournalTitle{The Journal of
  Physical Chemistry Letters}} \textbf{5}, 3662--3668 (2014).

\bibitem{Nemec2015}
H.~N{\v{e}}mec, V.~Zajac, P.~Ku{\v{z}}el, P.~Mal{\'{y}}, S.~Gutsch, D.~Hiller,
  and M.~Zacharias, \enquote{{Charge transport in silicon nanocrystal
  superlattices in the terahertz regime},} {\protect\JournalTitle{Physical
  Review B}} \textbf{91}, 195443 (2015).

\bibitem{Beaudoin2014}
A.~Beaudoin, B.~Salem, T.~Baron, P.~Gentile, and D.~Morris, \enquote{{Impact of
  n-type doping on the carrier dynamics of silicon nanowires studied using
  optical-pump terahertz-probe spectroscopy},} {\protect\JournalTitle{Physical
  Review B}} \textbf{89}, 115316 (2014).

\bibitem{Lu2013}
J.~Lu, H.~Liu, S.~X. Lim, S.~H. Tang, C.~H. Sow, and X.~Zhang,
  \enquote{{Transient Photoconductivity of Ternary CdSSe Nanobelts As Measured
  by Time-Resolved Terahertz Spectroscopy},} {\protect\JournalTitle{The Journal
  of Physical Chemistry C}} \textbf{117}, 12379--12384 (2013).

\bibitem{Tang2012}
H.~Tang, L.-G. Zhu, L.~Zhao, X.~Zhang, J.~Shan, and S.-T. Lee,
  \enquote{{Carrier Dynamics in Si Nanowires Fabricated by Metal-Assisted
  Chemical Etching},} {\protect\JournalTitle{ACS Nano}} \textbf{6}, 7814--7819
  (2012).

\bibitem{Jnawali2015}
G.~Jnawali, Y.~Rao, J.~H. Beck, N.~Petrone, I.~Kymissis, J.~Hone, and T.~F.
  Heinz, \enquote{{Observation of Ground- and Excited-State Charge Transfer at
  the C 60 /Graphene Interface},} {\protect\JournalTitle{ACS Nano}} \textbf{9},
  7175--7185 (2015).

\bibitem{Pijpers2010}
J.~J.~H. Pijpers, R.~Koole, W.~H. Evers, A.~J. Houtepen, S.~Boehme, C.~{de
  Mello Doneg{\'{a}}}, D.~Vanmaekelbergh, and M.~Bonn, \enquote{{Spectroscopic
  Studies of Electron Injection in Quantum Dot Sensitized Mesoporous Oxide
  Films},} {\protect\JournalTitle{The Journal of Physical Chemistry C}}
  \textbf{114}, 18866--18873 (2010).

\bibitem{Xu2009}
X.~Xu, K.~Chuang, R.~J. Nicholas, M.~B. Johnston, and L.~M. Herz,
  \enquote{{Terahertz excitonic response of isolated single-walled carbon
  nanotubes},} {\protect\JournalTitle{Journal of Physical Chemistry C}}
  \textbf{113}, 18106--18109 (2009).

\bibitem{Zhang2017}
W.~Zhang, X.~Zeng, X.~Su, X.~Zou, P.-A. Mante, M.~T. Borgstr{\"{o}}m, and
  A.~Yartsev, \enquote{{Carrier Recombination Processes in Gallium Indium
  Phosphide Nanowires},} {\protect\JournalTitle{Nano Letters}} \textbf{17},
  4248--4254 (2017).

\bibitem{Petersen2017}
J.~C. Petersen, A.~Farahani, D.~G. Sahota, R.~Liang, and J.~S. Dodge,
  \enquote{{Transient terahertz photoconductivity of insulating cuprates},}
  {\protect\JournalTitle{Physical Review B}} \textbf{96}, 115133 (2017).

\bibitem{Hendry2005}
E.~Hendry, M.~Koeberg, J.~M. Schins, H.~K. Nienhuys, V.~Sundstr{\"{o}}m,
  L.~D.~A. Siebbeles, and M.~Bonn, \enquote{{Interchain effects in the
  ultrafast photophysics of a semiconducting polymer: THz time-domain
  spectroscopy of thin films and isolated chains in solution},}
  {\protect\JournalTitle{Physical Review B - Condensed Matter and Materials
  Physics}} \textbf{71}, 1--10 (2005).

\bibitem{Hendry2006a}
E.~Hendry, M.~Koeberg, B.~O'Regan, and M.~Bonn, \enquote{{Local field effects
  on electron transport in nanostructured TiO2 revealed by terahertz
  spectroscopy},} {\protect\JournalTitle{Nano Letters}} \textbf{6}, 755--759
  (2006).

\bibitem{Shan2003}
J.~Shan, F.~Wang, E.~Knoesel, M.~Bonn, and T.~F. Heinz, \enquote{{Measurement
  of the frequency-dependent conductivity in sapphire.}}
  {\protect\JournalTitle{Physical review letters}} \textbf{90}, 247401 (2003).

\bibitem{Knoesel2004}
E.~Knoesel, M.~Bonn, J.~Shan, F.~Wang, and T.~F. Heinz, \enquote{{Conductivity
  of solvated electrons in hexane investigated with terahertz time-domain
  spectroscopy},} {\protect\JournalTitle{Journal of Chemical Physics}}
  \textbf{121}, 394--404 (2004).

\bibitem{Dakovski2007}
G.~L. Dakovski, S.~Lan, C.~Xia, and J.~Shan, \enquote{{Terahertz Electric
  Polarizability of Excitons in PbSe and CdSe Quantum Dots},}
  {\protect\JournalTitle{The Journal of Physical Chemistry C}} \textbf{111},
  5904--5908 (2007).

\bibitem{Kunneman2013}
L.~T. Kunneman, M.~Zanella, L.~Manna, L.~D.~A. Siebbeles, and J.~M. Schins,
  \enquote{{Mobility and Spatial Distribution of Photoexcited Electrons in
  CdSe/CdS Nanorods},} {\protect\JournalTitle{The Journal of Physical Chemistry
  C}} \textbf{117}, 3146--3151 (2013).

\bibitem{Bergren2014}
M.~R. Bergren, C.~E. Kendrick, N.~R. Neale, J.~M. Redwing, R.~T. Collins, T.~E.
  Furtak, and M.~C. Beard, \enquote{{Ultrafast Electrical Measurements of
  Isolated Silicon Nanowires and Nanocrystals},} {\protect\JournalTitle{The
  Journal of Physical Chemistry Letters}} \textbf{5}, 2050--2057 (2014).

\bibitem{Parkinson2012}
P.~Parkinson, C.~Dodson, H.~J. Joyce, K.~A. Bertness, N.~A. Sanford, L.~M.
  Herz, and M.~B. Johnston, \enquote{{Noncontact Measurement of Charge Carrier
  Lifetime and Mobility in GaN Nanowires},} {\protect\JournalTitle{Nano
  Letters}} \textbf{12}, 4600--4604 (2012).

\bibitem{Nemec2010}
H.~N{\v{e}}mec, P.~Ku{\v{z}}el, and V.~Sundstr{\"{o}}m, \enquote{{Charge
  transport in nanostructured materials for solar energy conversion studied by
  time-resolved terahertz spectroscopy},} {\protect\JournalTitle{Journal of
  Photochemistry and Photobiology A: Chemistry}} \textbf{215}, 123--139 (2010).

\bibitem{Nemec2009a}
H.~N{\v{e}}mec, H.-K. Nienhuys, E.~Perzon, F.~Zhang, O.~Ingan{\"{a}}s,
  P.~Ku{\v{z}}el, and V.~Sundstr{\"{o}}m, \enquote{{Ultrafast conductivity in a
  low-band-gap polyphenylene and fullerene blend studied by terahertz
  spectroscopy},} {\protect\JournalTitle{Physical Review B}} \textbf{79},
  245326 (2009).

\bibitem{Fekete2009}
L.~Fekete, P.~Ku{\v{z}}el, H.~N{\v{e}}mec, F.~Kadlec, A.~Dejneka,
  J.~Stuchl{\'{i}}k, and A.~Fejfar, \enquote{{Ultrafast carrier dynamics in
  microcrystalline silicon probed by time-resolved terahertz spectroscopy},}
  {\protect\JournalTitle{Physical Review B}} \textbf{79}, 115306 (2009).

\bibitem{DAngelo2016}
F.~D'Angelo, H.~N{\v{e}}mec, S.~H. Parekh, P.~Ku{\v{z}}el, M.~Bonn, and
  D.~Turchinovich, \enquote{{Self-referenced ultra-broadband transient
  terahertz spectroscopy using air-photonics},} {\protect\JournalTitle{Optics
  Express}} \textbf{24}, 10157 (2016).

\bibitem{Krewer2018}
K.~L. Krewer, Z.~Mics, J.~Arabski, G.~Schmerber, E.~Beaurepaire, M.~Bonn, and
  D.~Turchinovich, \enquote{{Accurate terahertz spectroscopy of supported thin
  films by precise substrate thickness correction},}
  {\protect\JournalTitle{Optics Letters}} \textbf{43}, 447 (2018).

\bibitem{Pedrotti2006}
F.~L. Pedrotti, L.~M. Pedrotti, and L.~S. Pedrotti, \emph{Introduction to
  Optics: International Edition} (Pearson, 2006), 3rd ed.

\bibitem{Kuzel2014a}
P.~Ku{\v{z}}el and H.~N{\v{e}}mec, \enquote{{Terahertz conductivity in
  nanoscaled systems: effective medium theory aspects},}
  {\protect\JournalTitle{Journal of Physics D: Applied Physics}} \textbf{47},
  374005 (2014).

\bibitem{Naftaly2007}
M.~Naftaly and R.~E. Miles, \enquote{{Terahertz time-domain spectroscopy for
  material characterization},} {\protect\JournalTitle{Proceedings of the IEEE}}
  \textbf{95}, 1658--1665 (2007).

\bibitem{Nashima2001}
S.~Nashima, O.~Morikawa, K.~Takata, and M.~Hangyo, \enquote{{Measurement of
  optical properties of highly doped silicon by terahertz time domain
  reflection spectroscopy},} {\protect\JournalTitle{Applied Physics Letters}}
  \textbf{79}, 3923--3925 (2001).

\bibitem{Karlsen2018}
P.~Karlsen, M.~V. Shuba, P.~P. Kuzhir, A.~G. Nasibulin, P.~Lamberti, and
  E.~Hendry, \enquote{{Sign inversion in the terahertz photoconductivity of
  single-walled carbon nanotube films},} {\protect\JournalTitle{arXiv}}
  \textbf{[cond-mat.mes-hall]}, 1804.11113 (2018).

\bibitem{Karlsen2017}
P.~Karlsen, M.~V. Shuba, C.~Beckerleg, D.~I. Yuko, P.~P. Kuzhir, S.~A.
  Maksimenko, V.~Ksenevich, H.~Viet, A.~G. Nasibulin, R.~Tenne, and E.~Hendry,
  \enquote{{Influence of nanotube length and density on the plasmonic terahertz
  response of single-walled carbon nanotubes},} {\protect\JournalTitle{Journal
  of Physics D: Applied Physics}} \textbf{51}, 014003 (2018).

\bibitem{Slepyan2010a}
G.~Y. Slepyan, M.~V. Shuba, S.~a. Maksimenko, C.~Thomsen, and a.~Lakhtakia,
  \enquote{{Terahertz conductivity peak in composite materials containing
  carbon nanotubes: Theory and interpretation of experiment},}
  {\protect\JournalTitle{Physical Review B}} \textbf{81}, 205423 (2010).

\bibitem{Byrnes2016}
S.~J. Byrnes, \enquote{{Multilayer optical calculations},}
  {\protect\JournalTitle{arXiv:1603.02720}}  (2016).

\end{thebibliography}
\end{document}